\documentclass{iopart}

\usepackage{graphicx} 
\usepackage{color}
\definecolor{Green}{rgb}{0,0.5,0}

\usepackage[english]{babel}
\usepackage[utf8]{inputenc}

\usepackage{amssymb}

\usepackage{textcomp}

\begin{document}

\title[Electronic triple-dot transport through a bilayer graphene island]{Electronic triple-dot transport through a bilayer graphene island with ultrasmall constrictions}
\author{D. Bischoff, A. Varlet, P. Simonet, T. Ihn, K. Ensslin}
\address{Solid State Physics Laboratory, ETH Zurich, 8093 Zurich, Switzerland}
\ead{dominikb@phys.ethz.ch}


\begin{abstract}
A quantum dot has been etched in bilayer graphene connected by two small constrictions to the leads. We show that this structure does not behave like a single quantum dot but consists of at least three sites of localized charge in series. The high symmetry and electrical stability of the device allowed us to triangulate the positions of the different sites of localized charge and find that one site is located in the island and one in each of the constrictions. Nevertheless we measure many consecutive single non-overlapping Coulomb-diamonds in series. In order to describe these findings, we treat the system as a strongly coupled serial triple quantum dot. We find that the non-overlapping Coulomb diamonds arise due to higher order cotunneling through the outer dots located in the constrictions. We extract all relevant capacitances, simulate the measured data with a capacitance model and discuss its implications on electrical transport.
\end{abstract}


\pacs{71.15.Mb, 81.05.ue, 72.80.Vp}

\maketitle

\section{Introduction}

Since becoming experimentally available in 2004~\cite{Novoselov2004}, graphene has triggered a wide range of research due to its many special properties. Graphene was early on suggested for building spin qubits~\cite{Trauzettel2007} as long spin coherence times are expected: In nature, predominantly the $^{12}$C isotope occurs which doesn't have a nuclear spin. Furthermore carbon has a low atomic number which should lead to small spin orbit coupling. Both effects are known to limit spin coherence times in for example GaAs~\cite{Meunier2007,Bluhm2010}.

In the past, graphene single electron transistors and quantum dots were fabricated in different ways: Structures were etched by reactive ion etching~\cite{Guettinger2008a,Guettinger2008b,Ponomarenko2008,Stampfer2008a,Stampfer2008b,Guettinger2009a,Guettinger2009b,Molitor2009,Moriyama2009,Moser2009,Schnez2009,Guettinger2010,Liu2010,Molitor2010,Schnez2010,Guettinger2011a,Guettinger2011b,Volk2011,Wang2011a,Droescher2012,Jacobsen2012,Mueller2012,Wang2012,Volk2013}, were defined by gates~\cite{Liu2010,Goosens2012,Allen2012} or alternative methods~\cite{Wang2011b,Bunch2005,Moser2009,Neubeck2010}. Single constrictions showing Coulomb-blockade were fabricated (for a list of references see e.g. Ref.~\cite{Bischoff2012}), single dots were fabricated~\cite{Guettinger2008a,Guettinger2008b,Ponomarenko2008,Stampfer2008a,Stampfer2008b,Guettinger2009a,Guettinger2009b,Moriyama2009,Moser2009,Schnez2009,Guettinger2010,Neubeck2010,Schnez2010,Guettinger2011a,Guettinger2011b,Wang2011b,Droescher2012,Jacobsen2012,Mueller2012,Goosens2012,Allen2012,Volk2013} and double dot systems were investigated~\cite{Wang2012,Wang2011a,Volk2011,Liu2010,Molitor2010,Molitor2009}. These structures were fabricated from single layer graphene~\cite{Guettinger2008a,Ponomarenko2008,Stampfer2008a,Stampfer2008b,Guettinger2009a,Guettinger2009b,Molitor2009,Moser2009,Schnez2009,Guettinger2010,Liu2010,Molitor2010,Neubeck2010,Schnez2010,Guettinger2011a,Guettinger2011b,Wang2011a,Wang2011b,Jacobsen2012,Mueller2012,Wang2012,Volk2013}, bilayer graphene~\cite{Volk2011,Wang2011a,Droescher2012,Goosens2012,Allen2012} or multilayer graphene~\cite{Bunch2005,Guettinger2008b,Moriyama2009}.

Compared to electrostatically defined GaAs nanostructures, etched graphene devices are strongly influenced by disorder which limits both the control as well as the reproducibility of the performed experiments. Desired properties as for example tunneling rates that are monotonically tunable by gate voltage or spin-blockade are so far not observed. Other expected effects as for example the observation of excited electronic states, reliable identification of few electron and hole states, shell filling or Kondo effect are only rarely observed and hard to reproduce.

In this paper we show electronic transport measurements on a bilayer graphene structure etched in a dot-shape with two attached leads. We try to avoid spurious effects of long constrictions by optimizing the design such that they are as short as possible. The constrictions are designed as thin as possible in order to achieve low tunneling rates and reduce the coupling of the island to the leads. Further the island is designed to be as small as possible in order to reach large charging and addition energies and we choose a high symmetry in order to simplify and improve the analysis of the measurements. We observe many consecutive closing and non-overlapping Coulomb diamonds as well as side-gate versus side-gate sweeps where only parallel resonances are visible. These findings are compatible with the system being a single-dot. At the same time we perform additional measurements which clearly show that in the same regime the system consists of three sites of localized charge in series. In order to explain these results, we model our structure as a serial triple dot system and estimate the relevant capacitances. Within this framework, the appearance of non-overlapping Coulomb-diamonds is caused by cotunneling through quantum dots located in the constrictions. These findings may explain why it is hard to reliably observe various of the above mentioned and expected effects in such a multi-dot system as presented in this paper.

\section{Experimental methods}

In the following we present electrical measurements of the device shown in Fig.~1a (fabrication similar to Ref.~\cite{Guettinger2009a}). All measurements were recorded at a temperature of 1.3 Kelvin with a symmetric dc-bias of 0.5~mV and all four side gates as well as the back-gate are always set to 0V unless stated differently. In the investigated regime no leakage currents were observed from any of the gates to any other part of the device. In order to check if the device was stable over the time of measurement, the first plot was recorded again after finishing with the measurements and found to be unchanged within experimental resolution.

\section{Results and discussion}

\begin{figure}[tbp]
	\centering\includegraphics[width=\textwidth]{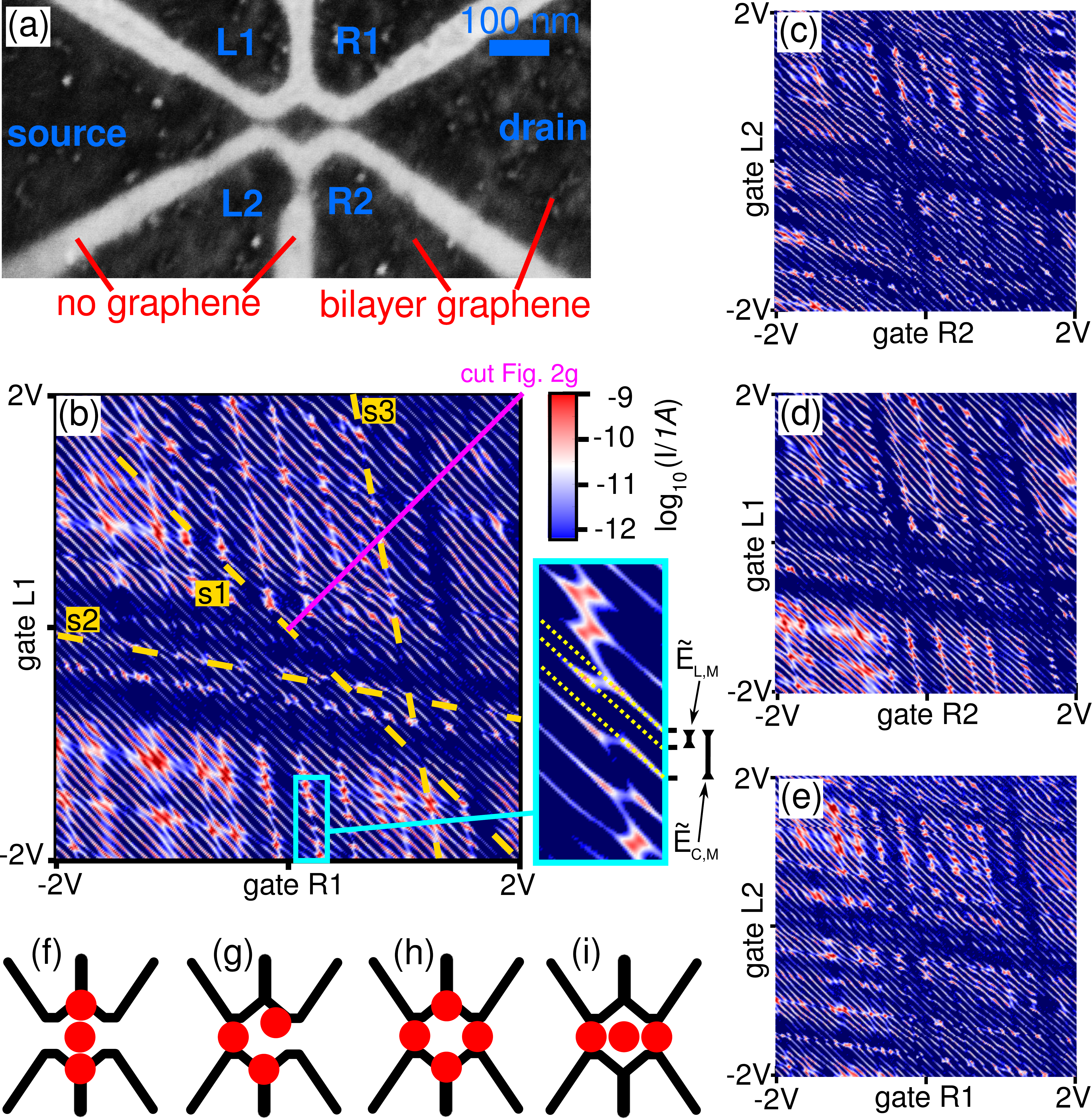}
	\caption{(color online) (a) Scanning electron microscopy image of the investigated device. (b) Measured current between source and drain (0.5~mV bias) as a function of the voltages applied at gates R1 and L1. Cyan box: Zoom into the marked region. (c) Current as a function of side gates R2 vs L2, (d) R2 vs L1 and (e) R1 vs L2. (f-i) Exemplary situations of how the sites of localized charge could be distributed. Situation (f) is not compatible with the measurements as all three sites would have equal capacitances to for example gates L1 and R1 and only slope -1 should then be visible in the corresponding measurement. Situation (g) is also not likely as different slopes would be expected for different combinations of left and right side gates. Situations (h) and (i) are so far both possible. Situation (h) will be excluded with further measurements shown in Figs.~2a,b.}
\end{figure}

Fig.~1b shows the current flowing from source to drain as a function of the gate voltages L1 and R1. Many diagonal lines (slope -1) are visible. These diagonal lines anti-cross with two other sets of lines. The three different negative slopes are marked with yellow dashed lines in Fig.~1b. Such lines with negative slopes are expected when a charge carrier is added to a localized site in the structure~\cite{Schroer2007,Wiel2003,Hofmann1995,Gaudreau2006,Rasmussen2008,Vidan2005}. The absolute value of the slope is determined by the ratio of capacitances of the two different gates to the given site of localized charge~\cite{Schroer2007,Wiel2003,Hofmann1995,Vidan2005,Gaudreau2006,Rasmussen2008}. The fact that a clear avoided crossing between the lines with different slopes is visible indicates that the different sites of localized charge are capacitively coupled to each other~\cite{Schroer2007,Wiel2003,Hofmann1995,Gaudreau2006,Rasmussen2008}. The observation of the three different sets of slopes suggests that in the investigated device at least three sites of localized charge are present that are capacitively coupled to each other. 

In Figs.~1c-e similar configurations as in Fig.~1b are shown: In all plots, one of the right side gates (R1,R2) is swept against one of the left side gates (L1,L2). In each of the four plots, there are three sets of lines that anti-cross with each other. Also the spacings between those lines as well as the slopes are comparable. The slopes are found to be s1=-1, s2=-0.19 and s3=-5.3 $\approx 1/\mathrm{s2}$. Errors from the data analysis and actual deviations between different lines in the same measurement are estimated to be smaller than 20\%.  Figs.~1f-i show exemplarily four situations of how three or more sites of localized charge might be distributed in the structure. From the measurements shown in Figs.~1b-e, the situations depicted in Figs.~1f,g can directly be excluded as they do not satisfy the symmetry observed in the experiment.

To obtain more information about the arrangement of the sites of localized charge, Figs.~2a,b show the current for the gate sweeps L1 vs L2 and R1 vs R2. Instead of three slopes, only one slope of -1 is visible. This indicates that each site of localized charge has the same capacitance to each of the left (right) gates and due to the structure's symmetry this signifies that all sites of localized charge must lie on a horizontal line going through the center of the structure. Together with the observation that there is always a slope of -1 in Figs.~1b-e, this  implies that one site of localized charge must necessarily be at the center of the structure (i.e. the situation depicted in Fig.~1h is not compatible with the measurements).  Figs.~2a,b also show clearly that the number of slopes in one measurement only yield a lower bound for the number of sites of localized charge involved. 

Additional details about the sites of localized charge are obtained by sweeping each side gate separately against the global back-gate as shown in Figs.~2c-f. In all four plots, three different sets of slopes are found: -2.2, -1.1, -0.36. Estimated errors are again below 20\%. From the voltage spacing along the side gate, the slope of -1.1 is identified to belong to the middle site of localized charge. 

In the following we estimate values for the involved capacitances by describing the system as a serial triple quantum dot~\cite{Waugh1995,Gaudreau2006,Schroer2007,Rasmussen2008,Hofmann1995,Vidan2005} employing the capacitive model for triple dots presented in Ref.~\cite{Schroer2007}. We name the three sites of localized charge ``dot left (L)'', ``dot middle (M)'' and ``dot right (R)''. In Fig.~1b, the spacing of lines belonging to the middle dot along axis L1 is about $\Delta L1_M\!\approx\!0.12\,\mathrm{V}$. This converts into a capacitance by: $C_{L1,M}\!=\!e/\Delta L1_M\!\approx\!1.3\,\mathrm{aF}$. As argued before, the capacitance of the middle dot is about the same to all four side gates. By the same procedure, one finds: $C_{Li,L}\!\approx\!C_{Ri,R}\!\approx\!0.5\,\mathrm{aF}$, $C_{Li,R}\!\approx\!C_{Ri,L}\!\approx\!0.1\,\mathrm{aF}$, $C_{BG,M}\!\approx\!1.2\,\mathrm{aF}$ and $C_{BG,L}\!\approx\!C_{BG,R}\!\approx 0.23\,\mathrm{aF}$ (where $i=1,2$). The values for the left and right dot have large errors due to the irregular spacing of the lines in the measurements. Via the plate capacitor model (285 nm silicon dioxide), the capacitances relative to the back-gate can be converted into areas: $A_{M}\!\approx\!(100\,\mathrm{nm})^2$ and $A_{L}\!\approx\!A_{R}\!\approx\! (40\,\mathrm{nm})^2$. Taking into account that stray field lines exist, these areas are compatible with the picture that the outer dots sit in the constrictions.

\begin{figure}[tbp]
	\centering\includegraphics[width=0.9\textwidth]{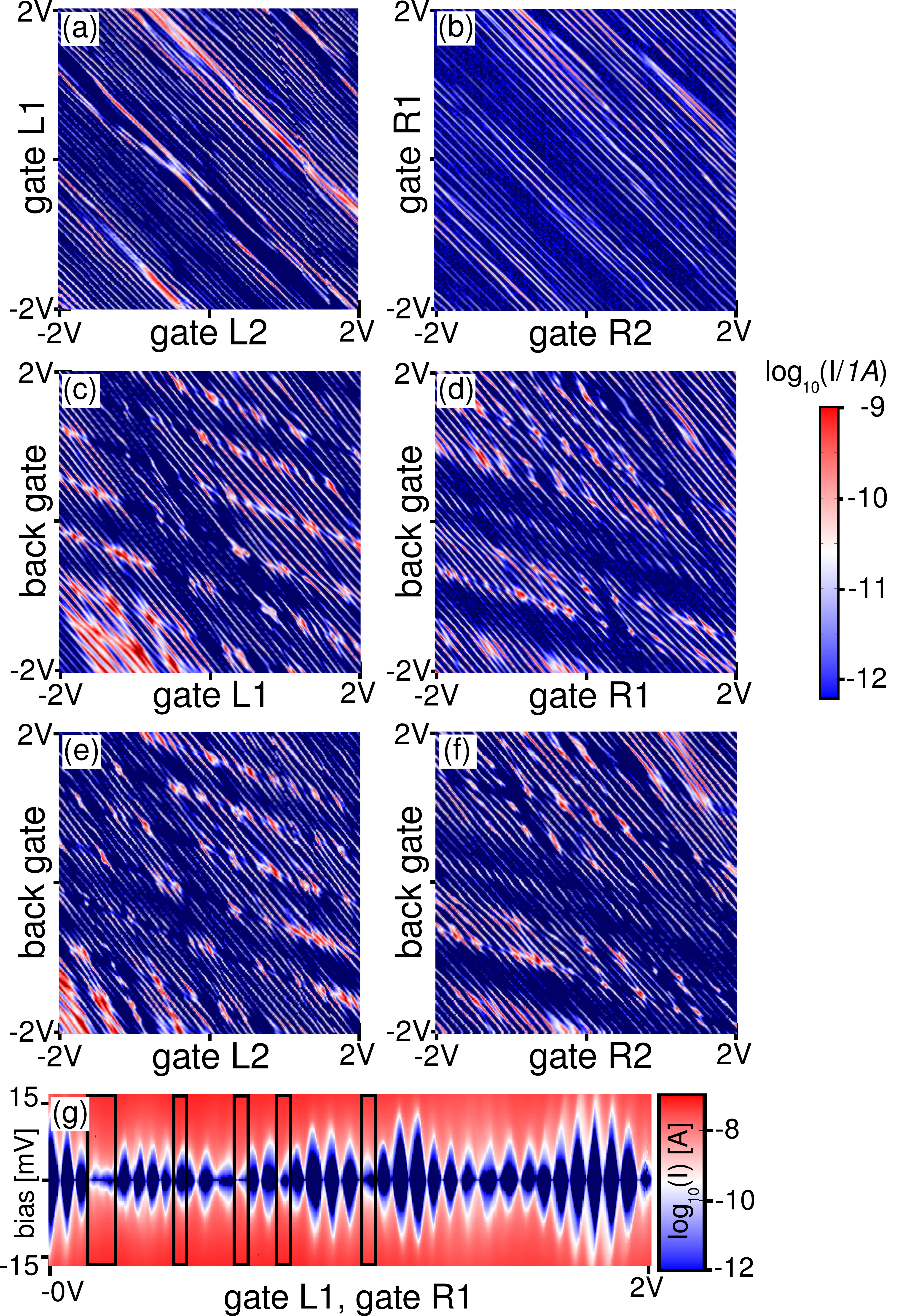}
	\caption{(color online) (a) Current flowing from source to drain (0.5 mV bias) as a function of gate voltages L2 vs L1, (b) R2 vs R1, (c) L1 vs back-gate, (d) R1 vs back-gate, (e) L2 vs back-gate and (f) R2 vs back-gate. (g) Current as function of applied bias voltage and gate voltages L1, R1 (both voltages are stepped simultaneously). The black frames mark regions where an electron is added to one of the outer dots.}
\end{figure}

Fig.~2g shows bias spectroscopy along a line in Fig.~1b going from the center to the upper right corner (see purple line in Fig.~1b). In the general case of three dots in series one could expect overlapping Coulomb-diamonds as it is unlikely that all three dot levels will be aligned with the leads for all the observed resonances. The measurement however shows mostly non-overlapping diamonds that close at zero bias. We will interpret this behavior at a later point. By comparing Fig.~1b with Fig.~2g, we can identify the diamonds marked with black rectangles to belong to situations where a charge is added to the outer dots. These are the diamonds that are significantly smaller in size than the others. The unmarked diamonds belong to charging events of the middle dot. Charging energies of the middle dot are therefore around 10 meV.

Capacitances that are so far unknown are the coupling capacitances between dots ($C_{L,M}$, $C_{M,R}$, $C_{L,R}$) and between each dot and the leads. If the system was a single dot system, the charging energy $E_C$ could be easily converted into the self capacitance $C^\Sigma$ (i.e. the sum of all capacitances from the dot to other objects) by: $E_C\!=\!e^2/C^\Sigma$. In a triple-dot system, this formula is more complicated as shown for example in Ref.~\cite{Schroer2007}. In the limit where the system is symmetric relative to the middle dot ($C_{L,M} = C_{M,R}$, $C^\Sigma_L = C^\Sigma_R$, ...) and if zero coupling from the left to the right dot is assumed ($C_{L,R}=0$), the formulae simplify to:
\begin{eqnarray}
E_{C,M} = \frac{e^2}{C^\Sigma_{M}} \cdot \frac{C^\Sigma_{L}C^\Sigma_{M}}{C^\Sigma_{L}C^\Sigma_{M} -2(C_{L,M})^2}\\
E_{C,L} = E_{C,R} = \frac{e^2}{C_{L}^\Sigma} \cdot \frac{C^\Sigma_{L}C^\Sigma_{M}-(C_{L,M})^2}{C^\Sigma_{L}C^\Sigma_{M}-2(C_{L,M})^2}
\end{eqnarray}
The charging energy is therefore mathematically given by the term from the single dot times a correction factor which is larger than one and which depends on the strength of the interdot coupling. The third value of interest is the electrostatic interdot coupling energy:
\begin{eqnarray}
E_{L,M} = E_{C,M} \cdot \frac{C_{L,M}}{C_{L}^\Sigma}
\end{eqnarray}
The ratio $E_{L,M}/E_{C,M}=\tilde{E}_{L,M}/\tilde{E}_{C,M}\!\approx\!1/3$ can be read off in the zoom left of Fig.~1b: The interdot charging energy is proportional to the voltage shift when an electron is added to the outer dot (see yellow dashed lines) and the voltage difference between charging lines of the middle dot is proportional to the charging energy of the middle dot. With this information and by assuming that the middle dot predominantly couples to the five gates and the two neighboring dots 
\begin{eqnarray}
C_M^\Sigma \approx C_{L,M} + C_{M,R} + C_{M,BG} + 4\times C_{M,\mathrm{side-gate}}
\end{eqnarray}
Eq.~1 can be solved to $C_{M}^\Sigma\!\approx\!21\,\mathrm{aF}$ and $C_{L,M}\!\approx\!7.2\,\mathrm{aF}$. Consequently the self capacitance of the outer dots is estimated to be $C_{L}^\Sigma\!=\!C_{R}^\Sigma\!\approx\!21\,\mathrm{aF}$ and the coupling between the outer dots and source/drain $C_{\mathrm{source},L}\!\approx\!C_{\mathrm{drain},R}\!\approx\!12\,\mathrm{aF}$ (assuming that the outer dots do not couple to the far lead). An overview of all estimated capacitances is shown in Tab.~1. These results show that a large part of the self capacitance arises due to coupling to neighboring dots and leads (compare also Ref.~\cite{Bischoff2012}).

\begin{table}[tbp]
\centering
\renewcommand{\arraystretch}{1.35}
\begin{tabular}{c | c c c}
			& dot L 				& dot M					& dot R					\\\hline
dot L			& ---					& $C_{L,M}\approx$ 7.2 aF		& $C_{L,R} =$ 0*			\\
dot M 			& $C_{L,M}\approx$ 7.2 aF 		& ---					& $C_{R,M}\approx$ 7.2 aF		\\
dot R			& $C_{L,R} =$ 0*			& $C_{R,M}\approx$ 7.2 aF	& ---					\\
source			& $C_{\mathrm{source},L}\approx$ 12 aF	& $C_{\mathrm{source},M}=$ 0*		& $C_{\mathrm{source},R}=$ 0*		\\
drain			& $C_{\mathrm{drain},L}=$ 0*		& $C_{\mathrm{drain},M} =$ 0*		& $C_{\mathrm{drain},R}\approx$ 12 aF	\\
back gate		& $C_{BG,L}\approx$ 0.2 aF		& $C_{BG,M}\approx$ 1.2 aF		& $C_{BG,R}\approx$ 0.2 aF		\\
side gate L$_i$		& $C_{Li,L}\approx$ 0.5 aF		& $C_{Li,M}\approx$ 1.3 aF		& $C_{Li,R}\approx$ 0.1 aF		\\
side gate R$_i$		& $C_{Ri,L}\approx$ 0.1 aF		& $C_{Ri,M}\approx$ 1.3 aF		& $C_{Ri,R}\approx$ 0.5 aF		\\\hline
self capacitance	& $C^\Sigma_L\approx$ 21 aF		& $C^\Sigma_M\approx$ 21 aF		& $C^\Sigma_R \approx$  21 aF		\\
\end{tabular}
\caption{Overview of the estimated capacitances between the three dots and different other parts of the device. Values marked by 0* are assumed to be zero as described in the text. Note that these values are to be taken as an order of magnitude rather than an exact value. Especially the coupling capacitances between neighboring dots and between dots and neighboring leads seem to vary strongly depending on the alignment of the different dot levels. This is for example seen in the modulation of the Coulomb diamond size in Fig. 2g.}
\end{table}

\begin{figure}[tbp]
	\centering\includegraphics[width=0.85\textwidth]{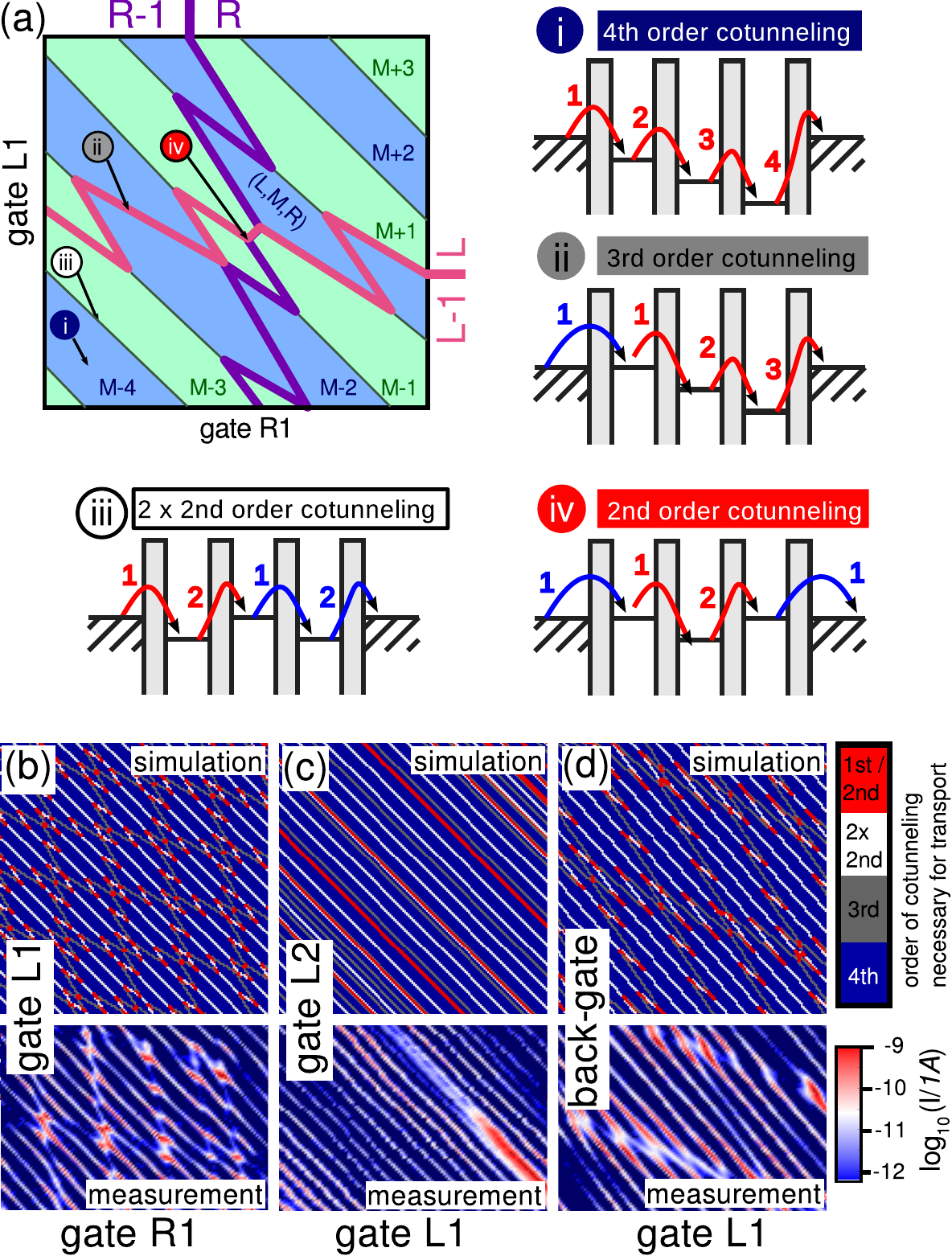}
	\caption{(color online) (a) Calculated charge stability diagram for gate R1 vs L1. Crossing the purple/salmon-colored lines changes the number of electrons by one in the right/left dot. Alternating green/blue areas mark constant numbers of electrons in the middle dot. Energy level diagrams (i)-(iv) are depicted schematically for four different situations. (b) Simulated charge stability diagram (upper part; colored according to order of cotunneling) versus measurement (lower part, current color scale as in Figs.~1,2) for gates R1 vs L1, (c) L1 vs L2 and (d) L1 vs back-gate. The voltage scale for the gates is the same in the simulation and the measurement.}
\end{figure}

In the following we use the extracted capacitances and the model from Ref.~\cite{Schroer2007} to calculate charge stability diagrams for different gate configurations. A part of such a stability diagram is shown in Fig.~3a where the integers L,M,R denote the number of charges in each dot. Alternating green and blue areas denote regions where the number of electrons in the middle dot is kept constant. By crossing the purple / salmon coloured lines, an additional charge carrier is loaded into the right / left dot. The alignment of the energy levels in the quantum dots relative to the leads are schematically depicted for four different situations (i-iv).

In order to compare these calculated charge stability diagrams with the plots from Figs.~1b-e,2a-f where current was measured, we color the charge stability diagrams according to a qualitative argument of how much current is expected to flow. In general, the higher the order of cotunneling that is at least necessary to transport a charge from source to drain, the lower is the current~\cite{Franceschi2001,Schroer2007,Rasmussen2008}. For each position in the charge stability diagram, the minimal order of cotunneling necessary for transport is determined and the corresponding position is color-coded according to the following rules (following the colorscale of Figs.~1,2 qualitatively): Resonant tunneling (very rare) and 2nd order cotunneling are marked in red, two times second order is white, third order is gray and fourth order is blue. Three such colored charge stability diagrams are shown in Figs.~3b-d (upper half) together with the corresponding measurement (lower half) for comparison. An overall similarity between these simulations and the measurements is observed. This model helps to understand why single Coulomb-diamonds are measured in Fig.~2g despite the serial-triple dot setting: As long as the dot levels in the outer two dots are not close to resonance with the leads, electrons can be loaded / unloaded into the middle dot via second order cotunneling through one of the outer dots.

While this qualitative model of a serial triple quantum dot describes the measurements well, there are many details that go beyond this model. For one, the spacing of the slopes corresponding to the outer dots varies quite significantly (see Figs.~1b-e,2a-f) suggesting that the area (and position) of the outer dots varies as a function of the gate voltages (compare also Ref.~\cite{Pascher2012}). Interestingly, the capacitance ratios to different gates stay about constant (see Figs.~1b-e,2a-f). Contrarily, the area of the middle dot doesn't seem to change much. Further the capacitance between the outer and the middle dots increases if the outer dots are close to resonance with the leads (decrease of the middle dot charging energy as seen in Fig.~2g). In general, the extracted capacitances should therefore be taken more as an estimate than an exact and constant value. Also in all measurements (Figs.~1b-e,2a-f) there are dark blue stripes where current is suppressed. This suggests a change in the coupling between the different dots and the leads or the formation of an additional site of localized charge in one of the leads. An additional limitation of the model is that various quantum effects as for example quantum confinement are neglected. Quantum confinement might play a role in the rather small dots in the constrictions such that the extracted charging energy would then correspond to an addition energy. Alternatively to the description chosen here, the system could be described as a coherently coupled three-site quantum dot molecule~\cite{Gustavsson2008}.

Despite all these variations, the overall appearance is surprisingly stable (there are about 70 electrons added to the center dot from the lower left corner to the upper right corner in Fig.~1b). This is in contrast to graphene nanoribbons where Coulomb diamonds generally appear to be quite irregular (see e.g. Ref.~\cite{Bischoff2012}). Qualitatively similar looking conductance plots in the plane of two gate voltages as shown in Figs.~1b-e,2a-f of this paper were also found for single layer \cite{Stampfer2008a,Stampfer2008b,Guettinger2011b} as well as trilayer \cite{Guettinger2008b} graphene quantum dots and triple dot behavior was found in scanning gate measurements on a single layer graphene quantum dot with significantly longer constrictions~\cite{Schnez2010}.

Measurements as presented here were recorded for several different randomly chosen regimes in back-gate voltage for two different cooldowns. While details vary, the overall appearance is similar. We also measured a second device with a nearly identical geometry and found similar results but with a generally much more suppressed current. These results suggest that the observed multi-dot behavior seems to be quite general and is not an artifact of choosing a peculiar gate regime.


In summary, we have fabricated a bilayer graphene quantum dot where both the dot size and the constriction size are at the limit of our currently employed technology. Even for these short and narrow constrictions, single electron charging inside the constrictions plays an important role for transport. As the device was designed such that it exhibits a high symmetry, it was possible to triangulate the positions of the different sites of localized charge: One quantum dot is formed in the island and at least one quantum dot is formed in each of the constrictions. While the quantum dot in the island is rather stable in shape and position, the dots in the constrictions vary in size and position. We have shown that depending on which gates are swept, either one slope or three slopes are visible in plots of the conductance in the plane of two gate voltages. In this regime where three sites of localized charge are arranged in series, we measured many consecutive and non-overlapping Coulomb diamonds. We discussed this behavior within the framework of higher order co-tunneling processes. In a system as presented in this paper it might be hard to observe certain effects as for example single dot Kondo physics: The Kondo effect itself occurs in the cotunneling regime of a dot coupled to Fermi leads. If these leads happen to be dots themselves with a level being populated with a spin (up or down) electron, then the concept of a screening cloud can not easily be applied. In order to create better controllable single graphene quantum dots, we believe that alternative fabrication methods need to be investigated in more detail. One promising candidate are the split-gate bilayer structures presented in Refs.~\cite{Allen2012,Goosens2012} where the edges are defined electrostatically.

\section*{Acknowledgements}
Financial support by the National Center of Competence in Research on “Quantum Science and Technology“ (NCCR QSIT) funded by the Swiss National Science Foundation is gratefully acknowledged.


\end{document}